\documentclass[12pt,a4paper]{article}
\usepackage{a4,epsf,here}
\def\K{\hbox{\bf K}}
\def\E{\hbox{\bf E}}
\def\sn{{\rm sn}}
\def\dn{{\rm dn}}
\def\cn{{\rm cn}}

\def\En{{\rm E}_n}
\def\Fn{{\rm F}_n}
\def\Eps{{\cal E}}
\def\SA{S_{\rm A}}
\def\TA{T_{\rm A}}
\def\M{{\rm M}}
\def\MA{{\rm M}_{\rm A}}
\def\trMA{{\rm tr}\,\MA}
\def\trM{{\rm tr}\,\M}
\def\be{\begin{equation}}
\def\ee{\end{equation}}
\def\bea{\begin{eqnarray}}
\def\eea{\end{eqnarray}}
\def\eq#1{(\ref{#1})}
\def\bs{\bigskip}
\def\ms{\medskip}
\def\fig#1{figure \ref{#1}}
\def\tab#1{table \ref{#1}}
\def\etal{{\it et al}}
\def\eps{\epsilon}
\def\siml{\,\hbox{\kern.1em \lower.6ex \hbox{$\sim$} \kern-1.12em
          \raise.6ex \hbox{$<$} }}
\def\simg{\,\hbox{\kern.1em \lower.6ex \hbox{$\sim$} \kern-1.12em
          \raise.6ex \hbox{$>$} }}
\newcommand{\Figurebb}[9]{
\begin{figure}[H]
\leavevmode
\epsfysize=#7cm
\epsfbox[#2 #3 #4 #5]{#6}
\par
\parbox{#8cm}{
\caption[figure]{\renewcommand{\baselinestretch}{0.8} \small
                                           \hspace{-0.3truecm}#9}
\label{#1}}
\end{figure}
}
\newcommand{\Table}[4]{
\begin{table}[H]\begin{center}{#3}
\parbox{#2cm}{
\vspace{0.5cm}
\caption[table]{\renewcommand{\baselinestretch}{0.8} \small
                                           \hspace{-0.3truecm}#4}
\label{#1}}
\end{center}
\end{table}
}

\begin{document}

\baselineskip 14pt

\centerline{\bf \Large Analytical perturbative approach to periodic orbits}

\bs

\centerline{\bf \Large in the homogeneous quartic oscillator potential}

\bs
\bs
\bs

\centerline{\bf M Brack$^1$, S N Fedotkin$^{1,2}$, A G Magner$^{1,2}$ 
                and M Mehta$^{1,3}$}

\ms
\bs

{\small

\centerline{$^1$Institute for Theoretical Physics, University of
Regensburg, D-93040 Regensburg, Germany}

\centerline{$^2$Institute for Nuclear Research, 252028 Prospekt 
Nauki 47, Kiev-28, Ukraine}

\centerline{$^3$Harish-Chandra Research Institute, Chhatnag Road, Jhusi,
Allahabad, 211019 India}

\bs

\centerline{\today~~(v3)}

\bs

\noindent
{\bf \large Abstract}

\ms
\noindent
We present an analytical calculation of periodic orbits in the 
homogeneous quartic oscillator potential. Exploiting the properties of 
the periodic Lam{\'e} functions that describe the orbits bifurcated 
from the fundamental linear orbit in the vicinity of the bifurcation 
points, we use perturbation theory to obtain their evolution away from 
the bifurcation points. As an application, we derive an analytical 
semiclassical trace formula for the density of states in the separable 
case, using a uniform approximation for the pitchfork bifurcations 
occurring there, which allows for full semiclassical quantization. For 
the non-integrable situations, we show that the uniform contribution of 
the bifurcating period-one orbits to the coarse-grained density of states
competes with that of the shortest isolated orbits, but decreases with 
increasing chaoticity parameter $\alpha$.

\section{Introduction}

The homogeneous quartic oscillator potential $V(x,y)=a\,x^4+b\,y^4+
(\alpha/2)\,x^2y^2$ has been the object of both classical, semiclassical 
and quantum-mechanical studies \cite{q4ls,q4po,btu,erda}. Lakshminarayan 
\etal~\cite{lakh} have investigated the fixed points in Poincar{\'e} 
surfaces of section corresponding to orbits of period four and determined 
empirically some of their scaling properties. Due to the homogeneity of 
the potential in the coordinates, orbits at different energies are related 
to each other through a simple scaling of coordinates and momenta. We may 
therefore fix the energy $E$ at an arbitrary value. The nonlinearity 
parameter that regulates the dynamics is the parameter $\alpha$. The 
system possesses periodic straight-line orbits along both axes which 
undergo stability oscillations under variation of $\alpha$. Infinite 
sequences of new periodic orbits bifurcate from each of these straight-line 
orbits and their repetitions, leading to {\it almost} completely chaotic 
dynamics \cite{daru} in the limit $\alpha\to\infty$.

In this paper we specialize to the symmetric case $a=b=1/4$ in which the 
potential has C$_{4v}$ symmetry; this potential shall in the following be 
denoted as the Q4 potential:
\be
V_{Q4}(x,y)=\frac14\left(x^4+y^4\right)+\frac{\alpha}{2}\,x^2y^2\,. 
\label{q4xy}
\ee
The straight-line orbits along the $x$ and $y$ axes then obey identical
equations of motion;
we denote them as the A orbits. The dynamics of the Q4 potential \eq{q4xy} 
is invariant under the symmetry operation (cf \cite{erda}) 
$\alpha\to (3-\alpha)/(1+\alpha)$, which corresponds to a rotation 
in the $(x,y)$ plane about 45 degrees and a simultaneous stretching of 
coordinates and time by a factor $[2/(1+\alpha)]^{1/4}$. The limit 
$\alpha\to\infty$ therefore is equivalent to the limit 
$\alpha\to -1$. There are three values of $\alpha$ for which the
potential is integrable: 1) $\alpha=0$, giving separability in $x$ and $y$; 
2) $\alpha=1$, which is the fixed point of the above symmetry 
operation, giving the isotropic quartic oscillator $V(r)=\frac14\,r^4$ with 
$r^2=x^2+y^2$; and 3) $\alpha=3$, giving separability after rotation about 
45 degrees.

\Figurebb{q4trm}{-30}{55}{795}{490}{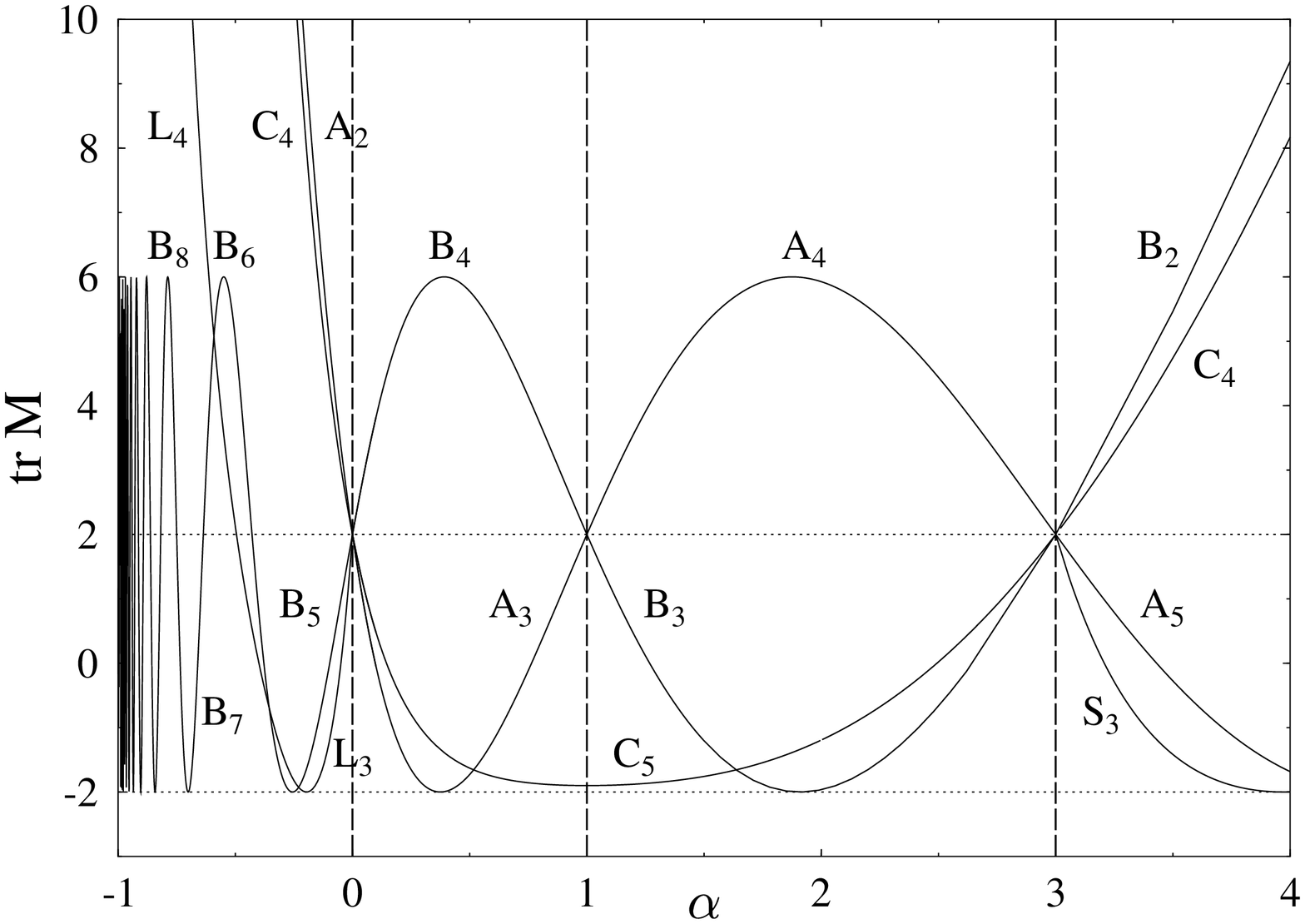}{8}{16.6}{
Stability discriminant $\trM$ of period-one orbits in the Q4 potential,
plotted versus $\alpha$. Shown are the curves for the primitive orbits
A, B and C, and the orbits L and S bifurcating from orbits A and B,
respectively. Subscripts indicate the Maslov indices $\sigma_{po}$ 
appearing in \eq{dgtf}. The vertical dashed lines at $\alpha=0$, 1 and 
3 correspond to the integrable situations.
}

In a recent paper \cite{lamp}, the period-one and period-two orbits 
bifurcating from the A orbits have been classified completely in terms of 
periodic Lam{\'e} functions. The motion of the primitive A orbit along the 
$y$ axis is given analytically by 
\be
x_A(t) = 0\,,\quad y_A(t) = y_0\,\cn(y_0t,\kappa)\,, 
         \qquad y_0=(4E)^{1/4}, \qquad \kappa^2=1/2\,,
\label{yaq4}
\ee
with the period $\TA = 4\K/y_0$, where $\K=K(\kappa)=F\left(\frac{\pi}{2},
\kappa\right)$ is the complete elliptic integral of the first kind with 
modulus $\kappa$, and $\cn(z,\kappa)$ is one of the Jacobi elliptic 
functions \cite{grry}. The turning points are $\pm y_0$. Note that this 
solution does not depend on the value of $\alpha$. The stability of the 
orbit A, however, does depend on $\alpha$. The linearized equation of 
motion in the transverse $x$ direction yields, after transformation to 
the scaled time variable $z=y_0t$, the Hill equation
\be
x''(z) + \alpha\,[1-\sn^2(z,\kappa)]\,x(z) = 0\,.
\label{lameq4}
\ee
This is a special case of the Lam{\'e} equation \cite{erde}
\be
\Lambda''(z) + \left[h - n(n+1)\,\kappa^2\,{\rm \sn}^2(z,\kappa)\right]\!
               \Lambda(z) 
       = 0 \qquad \hbox{with} \quad \kappa^2=\frac12\,, 
           \quad h = \alpha_n = \frac12 \,n\,(n+1)\,.
\label{lameq}
\ee
We therefore know here analytically the eigenvalues $h=\alpha_n$ of the 
Lam{\'e} equation, which correspond to the bifurcation points $\alpha_n$ 
of the A orbit \eq{yaq4}. This agrees with the analytical result for its 
stability discriminant, given by the trace of the stability matrix M, 
which has been derived long ago by Yoshida \cite{yosh}:
\be
\trMA = 4\,\cos\left(\frac{\pi}{2}\sqrt{1+8\alpha}\right) + 2\,.
\label{trmaq4}
\ee
It is easily seen that the bifurcation condition $\trMA=+2$ leads exactly 
to the values $\alpha_n$ in \eq{lameq}. 

In \fig{q4trm} we show the stability discriminant $\trM$ for some period-one 
orbits of the Q4 potential in the interval $-1\leq\alpha\leq 4$. The
integrable situations are indicated by the vertical dashed lines. The upper 
horizontal dotted line is the bifurcation line $\trM=+2$. The orbits B are
the two straight-line orbits along the diagonals $y=\pm x$, which are mapped 
onto the A orbits under the above-mentioned symmetry operation. Their motion 
is given by
\be
\pm\,x_B(t) = y_B(t) = y_0\,[2(1+\alpha)]^{-1/4}\,\cn(y_\alpha t,\kappa)\,, 
                       \qquad    
   y_\alpha = y_0\,[(1+\alpha)/2]^{1/4},
\ee
their period is $T_{\rm B}=\TA\,[2/(1+\alpha)]^{1/4}$, and their stability 
discriminant $\trM_{\rm B}$ is found from $\trMA$ in \eq{trmaq4} by 
replacing $\alpha\to (3-\alpha)/(1+\alpha)$. The orbit C is a rotational 
orbit which has a discrete degeneracy of two because of time reversal 
symmetry; its solutions $x_C(t)$ and $y_C(t)$ could not be found analytically
for arbitrary values of $\alpha$ (see section 3 for the case $\alpha=0$). 
L and S are the first librating orbits born from A and B, respectively, in 
isochronous (period-one) pitchfork bifurcations. Not shown are the period-one
orbits bifurcating from B for $\alpha\leq-3/7$. The results for $\trM$ for 
the C, L and S orbits were obtained numerically. The subscripts of all the 
orbits shown in \fig{q4trm} denote their Maslov indices $\sigma_{po}$ used 
in the semiclassical trace formula \eq{dgtf} in section 3.

The periodic solutions of the Lam{\'e} equation \eq{lameq} are the Lam{\'e} 
functions \cite{erde,inc1} Ec$_n^m(z,\kappa)$ and Es$_n^m(z,\kappa)$ which 
are even and odd functions of $z$, respectively, with $m$ zeros in 
$z\in[0,2\K)$. For integer $n$ they are polynomials of degree $n$ in the
Jacobi elliptic functions sn, cn and dn. Those with even $m$ have the 
period $T=2\K$, those with odd $m$ have $T=4\K$. For the special case 
$\kappa^2=1/2$, $m$ is fixed by $m={\rm int}[(n+1)/2]$ and there exists 
only one type of Lam{\'e} polynomial, Ec or Es, for each value of $n$. 
It is therefore sufficient here to denote these polynomials by E$_n(z)$ 
with $n=0,1,2,\dots$ Their explicit expressions up to $n=15$ have been 
given in \cite{lamp}. As there, we use from now on the short notation 
$\cn(z)=\cn(z,\kappa)$, $\sn(z)=\sn(z,\kappa)$ etc, keeping in mind that 
$\kappa^2=1/2$.

Nontrivial period doublings of the orbits A occur when $\trMA=-2$ (cf the 
lower horizontal line in \fig{q4trm}), which leads with \eq{trmaq4} to the 
critical values $\alpha_p = 2\,p\,(p+1)+3/8$ with $p = 0,1,2,\dots$ The 
corresponding solutions of \eq{lameq} with $n=(4p+1)/2$ are algebraic 
Lam{\'e} functions \cite{inc2,erd2} of period $8\K$ with $m=(2p+1)/2$. 
They are discussed in \cite{lamp} in connection with the period-two orbits 
born at the corresponding (island-chain type) bifurcations.

The purpose of this short paper is to demonstrate the use of the 
simple properties of the Lam{\'e} functions for analytical 
classical and semiclassical studies involving the periodic orbits
of the Q4 potential. In section 2 we shall formulate a perturbation
expansion for describing the evolution of the bifurcated orbits away 
from the bifurcation points $\alpha_n$, and in section 3 we apply some of
its results to semiclassical calculations of the density of states.

\section{Perturbation expansion around bifurcation points}

As shown in \cite{lamp}, the transverse motion of the orbits bifurcated 
from A is, infinitesimally close to the bifurcation values $\alpha_n$, 
exactly described by the Lam{\'e} functions. In \cite{lamp,mbgu} similar 
bifurcation cascades were investigated for H{\'e}non-Heiles type 
potentials. For these it was possible to determine the amplitude of the 
transverse motion of the bifurcated orbits from the conservation of the 
total energy, exploiting the asymptotic separability of these systems 
near the saddle energy into motions parallel and transverse to the 
bifurcating orbits. However, due to the scaling property of the Q4 
potential \eq{q4xy}, a variation of the energy does not affect the 
stability of the A orbits and hence the energy conservation cannot be 
exploited in the same way. In order to find the evolution of the new 
orbits away from their bifurcations, we propose here a perturbative 
series expansion of the equations of motion around the bifurcation points 
$\alpha_n$, leading to successive analytical expressions for the 
corrections to the simple (lowest-order) Lam{\'e} solutions.

The exact equations of motion for the Hamiltonian 
\be
H(p_x,p_y,x,y) = \frac12 \left(p_x^2+p_y^2\right) + V_{Q4}(x,y)
\label{ham}
\ee
with the Q4 potential \eq{q4xy} are, in the Newtonian form,
\be
\ddot x + x\,(x^2+\alpha\,y^2) = 0\,, \qquad
\ddot y + y\,(y^2+\alpha\,x^2) = 0\,.                    \label{eom}
\ee
In the following we expand the solutions of these equations around an
arbitrary bifurcation point $\alpha_n$ into Taylor series
\be
x(t) = x_A(t) + \epsilon\,x_1(t) + \epsilon^2 x_2(t) + \dots\,,
\qquad 
y(t) = y_A(t) + \epsilon\,y_1(t) + \epsilon^2 y_2(t) + \dots\,,
\label{exp}
\ee
whereby the small dimensionless expansion parameter $\eps$ is chosen, 
with support from numerical evidence, as
\be
\epsilon = \sqrt{|\alpha - \alpha_n|} \,.                \label{eps}
\ee
As zero-order solution, we have taken $x_A(t)=0$ and $y_A(t)=y_0\,{\rm 
E}_1(z)$ as given in \eq{yaq4} for the A orbit, since all bifurcated 
orbits are degenerate with the A orbit at the bifurcation points. 
Whereas this unperturbed solution is the same for all bifurcations, the 
corrections $x_k(t)$, $y_k(t)$ with $k=1,2,\dots$ will depend explicitly 
on the value $\alpha_n$ of the chosen bifurcation point. 

We now rewrite the exact equations of motion \eq{eom}, singling out the 
value of $\alpha_n$ and replacing the difference $\alpha-\alpha_n$ by 
$\epsilon^2$:
\be
\ddot x + \alpha_n\,x\,\,y^2 + \epsilon^2 x\,y^2 + x^3 = 0\,,\qquad
\ddot y + y^3 + \alpha_n\,y\,x^2 + \epsilon^2 y\,x^2 = 0\,.\label{eomp}
\ee 
Hereby we have chosen $\alpha\geq\alpha_n$ which holds for all bifurcated 
orbits with $n\geq 1$. The orbit L$_3$ ($n=0$) and the orbits F$_6$, P$_7$ 
($n=1/2$, see \cite{lamp}) bifurcate towards smaller values of $\alpha$, ie 
$\alpha\leq\alpha_0=0$ and $\alpha\leq\alpha_{1/2}=3/8$, respectively; for 
these cases the sign in front of $\epsilon^2$ in the equations \eq{eomp} 
must be reversed.

Next we insert \eq{exp} into \eq{eomp} and extract the equations obtained 
separately at each order in $\epsilon$. This leads to a recursive sequence 
of {\it linear} second-order differential equations in the scaled time 
variable $z$:
\be
x_k''(z) + \alpha_n\,\cn^2(z)\,x_k(z) = h_k(z)\,, \qquad
y_k''(z) + 3\,\cn^2(z)\,y_k(z)        = g_k(z)\,, \qquad
           k=1,2,3,\dots                                   \label{keq}
\ee
where the inhomogeneities on the rhs contain nonlinear combinations of 
$x_{k'}(t)$ and $y_{k'}(t)$ with $k'<k$. The homogeneous parts of these 
equations are identical to \eq{lameq4}, \eq{lameq} and have the periodic 
Lam{\'e} polynomials $\En(z)$ as solutions. According to their general 
theory \cite{erde,inc1} the second, linearly independent solutions are 
non-periodic for integer $n$; we shall in the following denote them by 
$\Fn(z)$. We normalize them such that their Wronskians with the $\En$ 
become unity: $W\!\left\{\En(z),\Fn(z)\right\}=1$. The solutions $\En$ 
and $\Fn$ are then related to each other by \cite{wiwa}
\be
\Fn(z) = \En(z) \int_0^z \frac{dz'}{[ \En(z')]^2}\,,   \qquad
\En(z) = \Fn(z) \int_z^0 \frac{dz'}{[ \Fn(z')]^2}\,.
\label{fsol}
\ee
In \tab{EnFn} we give the solutions $\En(z)$ and $\Fn(z)$ for the lowest
integer $n$. Hereby we have defined
\be
\Eps(z) = 2E(z)-z = \int_0^z \!\cn^2(u)\,du\,.
\label{Eps}
\ee
$E(z)$ is the incomplete elliptic integral of second kind, related
to that of the first kind $F(\varphi,\kappa)$ by
\be
E(z) = E(\varphi,\kappa)
     = \int_0^\varphi [1-\kappa^2\sin^2\theta]^{1/2}\,d\theta\,,\qquad 
   z = F(\varphi,\kappa)\,
     = \int_0^\varphi [1-\kappa^2\sin^2\theta]^{-1/2}\,d\theta\,. 
\ee
The function $\Eps(z)$ in \eq{Eps} is nonperiodic; its periodic part
is given, with $\E=E(\kappa)=E(\frac{\pi}{2},\kappa)$, by
\be
\mathrm{per}\{\Eps(z)\} = 2E(z) - z - \frac{\pi}{2\K^2}\,z
                        = 2E(z) - \frac{2\E}{\K}\,z\,.  \label{Epsper}
\ee
The last equality above follows from a known relation \cite{grry} 
between the complete elliptic integrals $\K$ and $\E$ for $\kappa^2=1/2$.

\Table{EnFn}{16.6}{
\begin{tabular}{|r|r|l|l|}
\hline
$n$ & $\alpha_n$ & E$_n(z)$ & F$_n(z)$ \\
\hline 
0 & 0 & E$_0(z)$ = Ec$_0^0(z)$ = \,1 & F$_0(z)$ = $z$ \\
1 & 1 & E$_1(z)$ = Ec$_1^1(z)$ = \,cn$(z)$                    
      & F$_1(z)$ = $2\,\sn(z)\,\dn(z) - \cn(z)\,\Eps(z)$ \\
2 & 3 & E$_2(z)$ = Es$_2^1(z)$ = \,dn$(z)$\,\sn$(z)$          
      & F$_2(z)$ = $z\,\dn(z)\,\sn(z) - \cn(z)$ \\
3 & 6 & E$_3(z)$ = Es$_3^2(z)$ = \,cn$(z)$\,dn$(z)$\,\sn$(z)$ 
      & F$_3(z)$ = $2 - 3\,\cn^4(z)-3\,\cn(z)\,\sn(z)\,\dn(z)\,\Eps(z)$\\
\hline
\end{tabular}
}{~The first four pairs of orthogonal solutions E$_n(z)$, F$_n(z)$ of 
the Lam{\'e} equation \eq{lameq}. We use the short notation
$\cn(z)=\cn(z,\kappa)$, $\sn(z)=\sn(z,\kappa)$ etc; here $\kappa^2=1/2$.
}

\vspace*{-0.5cm}

For half-integer $n$ -- appearing at the period-doubling bifurcations of 
the A orbits leading to the algebraic Lam{\'e} functions \cite{inc2,erd2} 
-- the linearly independent solutions $\Fn(z)$ are also periodic; the 
development given below must then be modified at some points. For 
simplicity, we limit ourselves in the following to the integer-$n$ cases 
occurring at the bifurcations of the primitive A orbits.

The general solutions of the equations \eq{keq} are of the standard form
\be
x_k(z) = c_k\,\En(z)  + d_k\,\Fn(z)  + H_k(z)\,,          \qquad
y_k(z) = a_k\,{\rm E}_2(z) + b_k\,{\rm F}_2(z) + G_k(z)\,,\label{ksol}
\ee
where the particular solutions of the inhomogeneous equations are given by
\bea
H_k(z) & = & \int_0^z h_k(z')\,[\,\Fn(z)\,\En(z')
                            - \Fn(z')\,\En(z)\,]\,dz'\nonumber\\
       & = & \En(z)\int_0^z du\, \frac{1}{[\En(u)]^2}
             \int_0^u h_k(w)\,\En(w)\,dw \;\;\;\;
\label{xkinh}
\eea
and
\bea
G_k(z) & = & \int_0^z g_k(z')\,[\,{\rm F}_2(z)\,{\rm E}_2(z')
                           - {\rm F}_2(z')\,{\rm E}_2(z)\,]\,dz'\nonumber\\
       & = & {\rm E}_2(z)\int_0^z du\, \frac{1}{[{\rm E}_2(u)]^2}
             \int_0^u g_k(w)\,{\rm E}_2(w)\,dw\,.\quad
\label{ykinh2}
\eea
The second parts of the above equations are useful for analytical
computations. The coefficients $a_k$, $b_k$, $c_k$ and $d_k$ in \eq{ksol} 
are determined recursively by requiring $x_k(z)$ and $y_k(z)$ to be 
periodic and to have the same symmetries as the lowest-order solutions 
$y_0(z)=y_A(z)$ and $x_1(z)=c_1\En(z)$, since these symmetries cannot be 
changed by varying $\alpha$ away from the bifurcations. The latter 
requirement leads immediately to $a_k=0$ for all $k$. Requiring $x_k(z)$ 
to be periodic allows us for $k\geq3$ to determine the constant $c_{k-2}$, 
appearing in different powers on the rhs of \eq{keq}, and (for any $k$) 
the constant $d_k$. Periodicity of $y_k(z)$ determines $b_k$ in terms of 
the $c_{k'}$ with $k'<k$. We find that $y(z)$ and $x(z)$ are overall even 
and odd functions of $\eps$, respectively, so that 
$c_{2k}=d_{2k}=b_{2k+1}=0$ and hence $x_{2k}(z)\equiv y_{2k+1}(z)\equiv 0$ 
for all $k=0,1,2,\dots$ Most of the integrals can be done analytically.

Up to this point, we have not respected the fact that the new bifurcated 
orbits develop their own periods which deviate from $\TA$ when moving 
away from the bifurcation points $\alpha_n$. The solutions $y_k(z)$ and 
$x_k(z)$ outlined above have, indeed, all the same period $\TA$ as the A 
orbit. A consequence of this is that these solutions do not conserve the
total energy $E$ as a function of $\alpha$. In the conventional perturbation 
theory, these unwanted effects are avoided by expanding the frequencies 
(or periods) of the perturbed system in powers of $\epsilon$ consistently 
along with the coordinates \eq{exp}. In our present case, this would have 
introduced more unknown parameters at each order, and the procedure of 
their determination would have become rather tedious. We have therefore
used an alternative approach by an {\it a posteriori} rescaling of the 
dimensionless argument $z=y_0t$ in the above solutions. At a given order 
of the perturbation expansion, we set $z=wt$ and determine the value of 
$w$ by the expansion of the new periods
\be
T = \frac{4\K}{w} = \TA \left[1 + \epsilon\,\tau_1 + \epsilon^2\tau_2 + 
                    \dots\right],                          \label{period}
\ee
leading to
\be
w = y_0\!\left[ 1 + \epsilon\,\tau_1 + \epsilon^2\,\tau_2 + \dots
    \right]^{-1}.                                             \label{wk}
\ee
The coefficients $\tau_k$ can be determined by writing the total energy 
in terms of the series \eq{exp} as
\be
E = \frac12\, ({\dot x}^2+{\dot y}^2)
  + \frac14\, (x^4+y^4) + \frac{\alpha}{2}\, x^2 y^2
  = E_0 + \epsilon\,E_1 + \epsilon^2\,E_2 + \dots\,,         \label{eq4}
\ee
inserting the above solutions for $x_k(wt)$ and $y_k(wt)$ with $w$ given
by \eq{wk}, and imposing that the energy $E$ be independent of $\epsilon$ 
(ie, of $\alpha$), which means that $E_0=E$ and $E_k=0$ for $k>0$. Whereas 
the lowest-order equation $E_0=E$ is trivially fulfilled for the solutions 
$x_A$, $y_A$ in \eq{yaq4}, the $E_k$ evaluated in terms of the above
solutions $x_k(y_0t)$, $y_k(y_0t)$ for $k>0$ are, indeed, not all equal to 
zero. However, in terms of the rescaled solutions $x_k(wt)$, $y_k(wt)$ we 
can impose $E_k=0$ for all $k>0$ successively by choosing the coefficients 
$\tau_k$ appropriately. An alternative way of deriving the same results 
(and thereby to check the algebra) is to consider the perturbation 
expansion \eq{eq4} with new ``variables'' (constants of the motion) $E_k$ 
which specify the total energy $E$. The energy conservation equation 
\eq{eq4} then leads at each order $k$ to a first-order differential 
equation which can be integrated analytically with respect to the 
perturbative solutions $y_k$, with the help of \eq{keq} for the $x_k$. 
From their periodicity conditions we obtain the $E_k$, and hence
the $\tau_k$ valid for all values of $\alpha$ and $E$. 
We found for all orbits investigated here that $\tau_1=\tau_2=\tau_3=0$ 
and $\tau_4\neq0$, so that $T$ varies with $\alpha$ like 
\be
T(\alpha) = \TA\!\left[1+\tau_4\,(\alpha-\alpha_n)^2+\dots\right],
\label{period4}
\ee 
whereby $\tau_4$ is an energy-independent constant. Table \ref{t4tab}
contains the values of $\tau_4$ of the first eight stable orbits born 
at the pitchfork bifurcations of A. They will be discussed further in 
section 3.2. We see that the L$_3$ orbit plays a special role; note that 
this orbit only exists at $\alpha\leq\alpha_0=0$, whereas all the other 
orbits given in the table exist only at $\alpha\geq\alpha_n$.

\Table{t4tab}{16.6}{
\begin{tabular}{|r|r|l|l|}
\hline
$n$ & $\alpha_n$ & O$_n$ & $\tau_4(n)$ \\
\hline 
0  & 0   & L$_3$    & $3\pi^2\!/16\K^4$ \\
3  & 6   & R$_5$    & $-$11/1260 \\
5  & 15  & L$_7$    & $-$475/354816 \\
7  & 28  & R$_9$    & $-$4807/12627300 \\
9  & 45  & L$_{11}$ & $-$160425/1092591808 \\
11 & 66  & R$_{13}$ & $-$75981/1115482060 \\
13 & 91  & L$_{15}$ & $-$4626964303/129311102954880 \\
15 & 120 & R$_{17}$ & $-$56892225/2767492640836 \\
\hline
\end{tabular}
}{~Lowest non-vanishing perturbation term $\tau_4$ in the period
\eq{period4} of the first eight stable orbits O$_n$ born at the 
bifurcations of the primitive orbit A. 
}

\vspace*{-0.5cm}

As an illustration of our method, we present the results for the orbit R$_5$ 
born at $\alpha_3=6$ and existing only for $\alpha\geq6$. We obtain the 
following analytical solutions up to order $\epsilon^4$ (with $z=wt$)
\bea
x_1(t) & = & c_1\,\sn(z)\,\dn(z)\,\cn(z)\,,\qquad
  x_3(t) = -\,c_1\,\sn(z)\,\dn(z)\,\cn(z)\left[\frac{25}{132} +
             \frac{7}{180}\,\cn^4(z)\right],\nonumber\\
y_0(t) & = & w\,\cn(z) \; = \; y_A(z)\,,\quad \qquad\! 
  y_2(t) = -\,\frac{c_1^2}{4w}\,\cn(z)\,[1+\cn^4(z)]\,,\nonumber\\
y_4(t) & = & \frac{c_1^2}{w}\,\cn(z)\left\{\frac{1}{10\,080}
               \left[-\,797-206\,\cn^4(z)+111\,\cn^8(z)\right]
               +\frac{5}{264}\left[1+\cn^4(z)\right]\right\},\label{r5sol}
\eea
where $c_1 = w\sqrt{11/45}$, and $\tau_4=-11/1260$ so that 
$w=y_0/[1-(11/1260)\eps^4]$. 

\newpage

\Figurebb{y4R5}{42}{50}{760}{515}{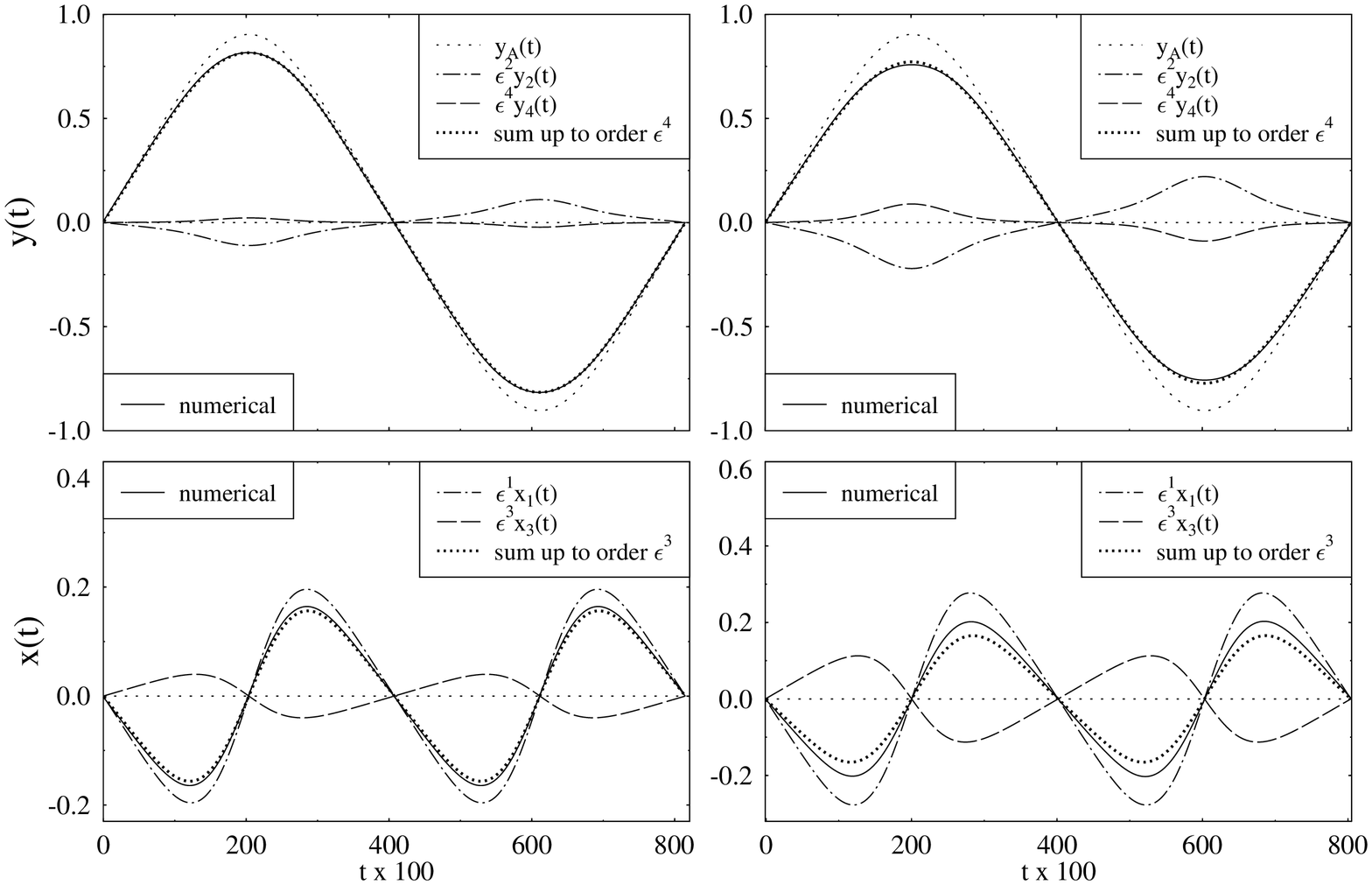}{10.7}{16.6}{
Coordinates $x(t)$ and $y(t)$ of the orbit R$_5$, evaluated at $\alpha=7$, 
ie $\epsilon=1$ (left side) and $\alpha=8$, ie $\epsilon=\sqrt{2}$ (right 
side). Solid lines: numerical results. Heavy dotted lines: perturbation 
series \eq{exp} including the terms in \eq{r5sol} up to 3rd and 4th order 
for $x(t)$ and $y(t)$, respectively. Other lines give partial 
contributions as labeled by the inserts. (Units such that $\hbar=1$; 
$E=1/6$.)
}

In figure \ref{y4R5} we compare the results obtained by summing the above
analytical results according to \eq{exp} up to order $\eps^4$, shown by 
the heavy dotted lines, with the results obtained by numerical solution 
of the exact equations of motion \eq{eom}, shown by the solid lines. The 
single contributions obtained at each order in $\eps$ are shown by the 
other lines, as labeled by the inserts. The left panels are calculated
for $\alpha=7$ where $\eps=1$, and the right panels for $\alpha=8$ where 
$\eps=\sqrt{2}$. The convergence is surprisingly good even when the 
expansion parameter $\eps$ is larger than unity. This is due to the 
rapidly decreasing amplitudes of the $x_k(z)$ and $y_k(z)$ with increasing 
$k$. Note that for $y(t)$, where the agreement is perfect for $\eps=1$ and 
still very good for $\eps=\sqrt{2}$, we have included three terms, whereas 
$x(t)$ only contains two terms. Obtaining $x_5(z)$ would have required to 
calculate both $y_6(z)$ and $x_7(z)$ and to make them periodic, which -- 
though analytically possible -- would have been rather cumbersome. But 
the fact that the remaining errors in $x(t)$ are of the same order as 
$\eps^4y_4(z)$ suggests that adding the third term $\eps^5x_5(z)$ would 
lead to an equally good convergence for $x(t)$. Similar results were 
also obtained for other bifurcated orbits. 

For the period-two orbits born at period-doubling bifurcations, the 
algebraic Lam{\'e} functions have to be used. The repeated integrations 
arising in the perturbation expansion then become more difficult and we 
could not do all of them analytically. Resorting to numerical 
integrations, however, whereby the coefficients $b_k$, $c_k$ and $d_k$ 
in \eq{ksol} were determined numerically by iteration, we could reach a 
similar convergence of the perturbation series.

\newpage

\section{Semiclassical trace formulae for the density of states}

In this section we shall apply our perturbative results to investigate
the role of the pitchfork bifurcations of the A orbit in semiclassical 
calculations of the density of states. We coarse-grain both the 
exact and the semiclassical density of states by a convolution with a 
normalized Gaussian $\exp\{-(E/\gamma)^2\}/\gamma\sqrt{\pi}$. The 
quantum-mechanical coarse-grained density of states is then given by 
\be
g_{qm}(E) = \frac{1}{\gamma\sqrt{\pi}}\sum_n \exp\{-[(E-E_n)/\gamma]^2\}  
\label{gqm}
\ee
in terms of the exact quantum spectrum $\{E_n\}$ which we have obtained 
by diagonalization of \eq{q4xy} in a harmonic oscillator basis. For 
larger values of $\gamma$, the prominent gross-shell structure in the 
density of states is emphasized while finer details of its oscillations 
are suppressed. In the limit $\gamma\to 0$ the sum of delta functions 
is recovered. The oscillating part of \eq{gqm} is defined as 
\be
\delta g_{qm}(E) = g_{qm}(E)-g_{TF}(E)\,,                  \label{dgqm}
\ee 
where $g_{TF}(E)$ is the average part obtained in the Thomas-Fermi 
approximation which for the Q4 potential can be calculated analytically:
\bea
g_{TF}(E)&=&\frac{2}{\pi\hbar^2}\,K\!\left(\!\sqrt{(1-\alpha)/2}\,\right)\!
            \sqrt{E}\hspace{3.6cm}\hbox{for }-1<\alpha\leq +1\,,\nonumber\\
         &=&\frac{2}{\pi\hbar^2}\,\sqrt{\frac{2}{\alpha+1}}\,
            K\!\left(\!\sqrt{(\alpha-1)/(\alpha+1)}\,\right)\!\sqrt{E}
            \qquad\quad\hbox{for }\;\alpha\geq +1\,.          \label{gtf}
\eea
Higher-order $\hbar$ corrections to the average density of states 
\cite{book} are negligible in the present system. 

The semiclassical trace formula for the density of states has the general 
form
\be
\delta g_{sc}(E) = \frac{1}{\pi\hbar}\sum_{po}{\cal A}_{po}(E)\,
                   f_\gamma(T_{po})\,
                   \cos\left[\frac{1}{\hbar}\,S_{po}(E)
                   -\sigma_{po}\frac{\pi}{2}\right].           \label{dgtf}
\ee
The sum goes over all periodic orbits ($po$) of the classical system. 
$S_{po}(E)$ are the action integrals $\oint {\bf p}\cdot d{\bf q}$ along 
the periodic orbits and $\sigma_{po}$ the so-called Maslov indices. The 
amplitudes ${\cal A}_{po}(E)$ depend on the number of constants of motion 
of the system. The factor $f_\gamma(T_{po})$ is given by \cite{book}
\be
f_\gamma(T_{po}) = \exp\left[{-(\gamma\,T_{po}/2\hbar)^2}\right],
\ee
where $T_{po}=dS_{po}(E)/dE$ are the periods of the orbits. This 
coarse-graining factor favours the contributions of the shortest
orbits to the gross-shell structure obtained with larger values of 
$\gamma$. When the system possesses no other constant of the motion
besides the energy, all orbits are isolated and Gutzwiller's original 
form \cite{gutz} of the amplitudes ${\cal A}_{po} (E)$ applies. In the 
presence of continuous symmetries, and for integrable systems in general, 
other forms must be used (see \cite{book} for a survey of trace formulae 
and the calculation of ${\cal A}_{po}$ and $\sigma_{po}$). 

In the presence of bifurcations, the Gutzwiller amplitudes ${\cal A}_{po}
(E)$ for isolated orbits cannot be used. However, the uniform 
approximation for generic pitchfork bifurcations \cite{ozha,ssun} can 
be applied to describe the isochronous bifurcations of the A orbit 
discussed in this paper. The factor 2 appearing in the period doubling 
for the generic case here plays the role of the extra degeneracy factor 
2 of the bifurcated orbit pairs, which is due either to their reflection 
symmetry at the symmetry axis containing the A orbit (for the librating
orbits L$_n$) or to the time reversal symmetry (for the rotating orbits 
R$_n$). Adapting the results of \cite{ssun} to the present system, the 
combined contribution to the semiclassical density of states from all 
the orbits participating in the bifurcation at $\alpha_n$ (including 
their degeneracy factors) is given by
\be
\delta g_{sc}^{bif}(E,\alpha_n) = \frac{\TA\,\Gamma(1/4)}
                {\pi\sqrt{2\pi}\,\hbar^{5/4}|a_n|^{1/4}k^{3/4}}\,
                \cos\left[\frac{k\SA}{\hbar}-k\nu_n\frac{\pi}{2}
                -(-1)^n\frac{\pi}{4}-{\rm sign}(a_n)\frac{\pi}{8}\right]. 
\label{levbif}
\ee
Here $k$ is the repetition number of the orbits and $\nu_n$ the Maslov 
index, corresponding to $\sigma_{po}$ in \eq{dgtf}, of the unstable 
primitive orbit involved. The parameter $a_n$ stems from the normal form 
of the action function $S(q',p)$ used in the phase-space representation 
of the trace integral at the $n$-th bifurcation of the A orbit. From the
expansions given in \cite{ssun} for the properties of the periodic orbits 
near the bifurcation, and using our results in the previous section, we 
can determine $a_n$ to be 
\be
a_n = \frac{1}{4S_{\rm A}}\left(\frac{4\pi}{2n+1}\right)^2
      \frac{1}{\tau_4(n)}\,.                            \label{an}
\ee

In the following, we want to examine the importance of the contribution 
\eq{levbif} in relation to that of other non-bifurcating orbits of the 
system. Quite generally, in an integrable two-dimensional system the 
leading families of degenerate orbits have an amplitude proportional to
$\hbar^{-3/2}$ (except, eg, for harmonic oscillators with rational frequency 
ratios, where the leading amplitudes go like $\hbar^{-2}$, see \cite{brja}). 
They are therefore expected to dominate over the bifurcating orbits which 
here have an amplitude proportional to $\hbar^{-5/4}$ as seen in \eq{levbif}. 
In section 3.1 below we will test the relative importance of the latter 
for the particular integrable situation given in our system for $\alpha=0$. 
On the other hand, the standard Gutzwiller amplitudes for isolated orbits 
in a non-integrable system go like $\hbar^{-1}$, so that the bifurcating 
orbits can be expected to play a larger relative role. As we will see in 
section 3.2 their relative weight is, however, subject to a rather subtle 
balance between the stability of the shortest isolated orbits and the 
$\alpha$ dependence of the factor $|a_n|^{-1/4}$ in the amplitude of 
\eq{levbif}. 

\subsection{The separable case $\alpha=0$}

For $\alpha=0$ the potential \eq{q4xy} is separable and thus defines
an integrable system in which the majority of the periodic orbits live 
on a 2D torus in phase space. The EBK quantization can therefore be 
applied to obtain a semiclassical quantum spectrum from which a trace 
formula is readily derived \cite{beta}. However, the system possesses  
also the isolated orbit A which undergoes a bifurcation at $\alpha=0$. 
As we see from \fig{q4trm} the orbit A, which is stable (A$_3$) for 
$0<\alpha<1$, becomes unstable (A$_2$) for $\alpha<0$; the orbit L$_3$ 
bifurcating from it exists only for $\alpha\leq0$ and is stable for 
$-0.5\siml\alpha<0$. This gives a nontrivial contribution to the 
density of states whose role will be investigated numerically below.

Straightforward EBK quantization of the 2D torus gives the approximate 
spectrum
\be
E^{\rm EBK}_{n_xn_y} = (1/4)\,(6\pi\hbar/4\K)^{4/3}        
                       \left[(n_x+1/2)^{4/3}+(n_y+1/2)^{4/3}\right].
                       \qquad (n_x,n_y=0,1,2,\dots)      \label{ewkb} 
\ee
From this we obtain in the standard way \cite{beta,crli} the following 
Berry-Tabor type trace formula to leading order in $\hbar$:
\be
\delta g_{sc}^{2D}(E) = \left(\frac{4\K}{2\pi\hbar}\right)^{\!3/2}\!
                         (4E)^{1/8}\; 2\!\!\! \sum_{n,m=1}^\infty(-1)^{n+m}\,
                         \frac{nm}{(n^4+m^4)^{5/8}}\,f_\gamma(T_{nm})\,
                         \cos\left[\frac{1}{\hbar}\,S_{nm}(E)
                                   -\frac{\pi}{4}\right]. \label{dgsc2}
\ee
The classical actions and periods
\be
S_{nm}(E)=(4\K/3)\,(4E)^{3/4}\,(n^4+m^4)^{1/4}, \qquad
T_{nm}(E)=4\K\,(4E)^{-1/4}\,(n^4+m^4)^{1/4}        \label{snm}
\ee
of the 2D rational tori ($n,m$) correspond to the periodic orbits
\be
x(t)=w_x\,\cn(w_xt+\phi)\,, \qquad y(t)=w_y\,\cn(w_yt)\,, \quad\;\,
\ee
with
\be
w_x=(4E)^{1/4}\,n\,(n^4+m^4)^{-1/4}, \qquad 
w_y=(4E)^{1/4}\,m\,(n^4+m^4)^{-1/4}.
\ee
These orbits form degenerate families described by the parameter
$\phi\in[0,4\K)$. The $(n,n)$ resonances are the families containing
the $n$-th repetitions of the orbits B (for $\phi=0$ or $2\K$) and C 
(for $\phi=\K$ or $3\K$). Summation over all resonances $(n,m)$ with 
$n,m>0$ yields the trace formula \eq{dgsc2}.

\newpage

However, the system also contains the isolated resonances $(n,0)$ and 
($0,m$) with $n,m=1,2,3,\dots$ which correspond to the (repeated) 
one-dimensional A orbits in the $x$ and $y$ direction. With the uniform
approximation \eq{levbif} discussed above, we can include them in the
trace formula, together with the orbits L$_3$ born at the bifurcation
A$_3\to$ A$_2$ + L$_3$. This gives the following common contribution of 
all A and L orbits and their $k$-th repetitions
\be
\delta g_{sc}^{\rm AL}(E)= \frac{(4\K)^{3/4}}{(\pi\hbar)^{5/4}}\,
                           (4E)^{-1/16} \sum_{k=1}^\infty (-1)^k
                           \frac{1}{k^{3/4}}\,f_\gamma(k\TA)\,
                           \cos\left[\frac{k}{\hbar}\SA(E)-\frac{3\pi}{8}
                           \right]\!,                       \label{dgsc1} 
\ee
where $\SA(E)=(4\K/3)\,(4E)^{3/4}$ is the action of the primitive A orbit. 
Hereby we have used the value $\tau_4=3\pi^2\!/16\K^4$ for the L$_3$ orbit 
as given in \tab{t4tab}, and the relation \cite{grry} 
$\Gamma(1/4)=(4\sqrt{\pi}\,\K)^{1/2}$.

The total oscillating part of the semiclassical density of states at 
$\alpha=0$ then is given by the sum of the contributions \eq{dgsc2} and 
\eq{dgsc1}. We observe that the latter is of order $\hbar^{1/4}$ relative 
to the leading-order contributions of the 2D torus families. (For isolated 
orbits, this relative factor would be $\hbar^{1/2}$ as mentioned above.) 
We therefore expect that the bifurcating orbits have a non-negligible 
influence on the density of states, at least for low energies where the 
negative power of $E$ in the amplitude of \eq{dgsc1} does not suppress 
their contribution too much. 

In \fig{dg_1} we show the coarse-grained level densities obtained with 
$\gamma=1$ (in units such that $\hbar=1$); the periodic orbit sums in 
\eq{dgsc2} and \eq{dgsc1} can here be limited to $n_{max}=m_{max}=
k_{max}=2$. In the upper panel, we show separately the semiclassical 
results \eq{dgsc2} of the 2D tori (solid line) and \eq{dgsc1} of the 
bifurcating orbits (dotted line). Both are seen to have a monotonously 
increasing amplitude of the oscillations, whereas the quantum result 
(dashed line) exhibits a pronounced beating structure. When adding both 
semiclassical contributions, as shown in the lower panel, the quantum 
result is nicely reproduced. The quantum beat is clearly the result of 
the interference between the shortest orbits of the torus family on one 
hand and the bifurcating isolated orbits A and L$_3$ on the other hand.

\Figurebb{dg_1}{10}{35}{795}{485}{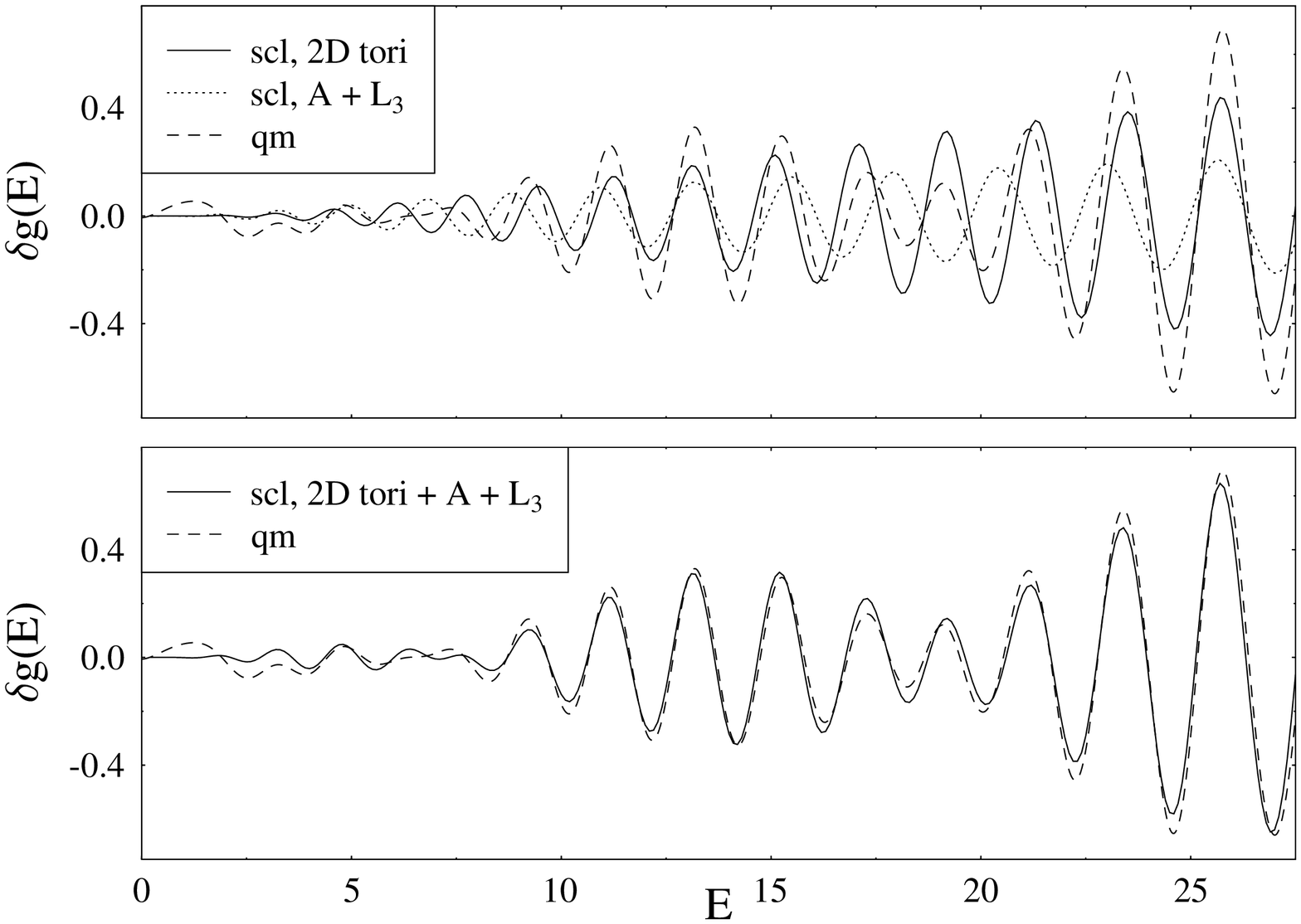}{9}{16.6}{
Oscillating part of the coarse-grained density of states 
at $\alpha=0$ (units such that $\hbar=1$; Gaussian smoothing width 
$\gamma=1$). Dashed lines: quantum-mechanical results \eq{dgqm}. 
Solid and dotted lines: semiclassical results with $n_{max}=m_{max}=
k_{max}=2$. Upper panel: separate contributions \eq{dgsc2} of 2D tori
and \eq{dgsc1} of bifurcating orbits A + L$_3$. Lower panel: the solid
line is here the sum of both semiclassical contributions.
}

\newpage

\Figurebb{dg_06}{20}{35}{795}{500}{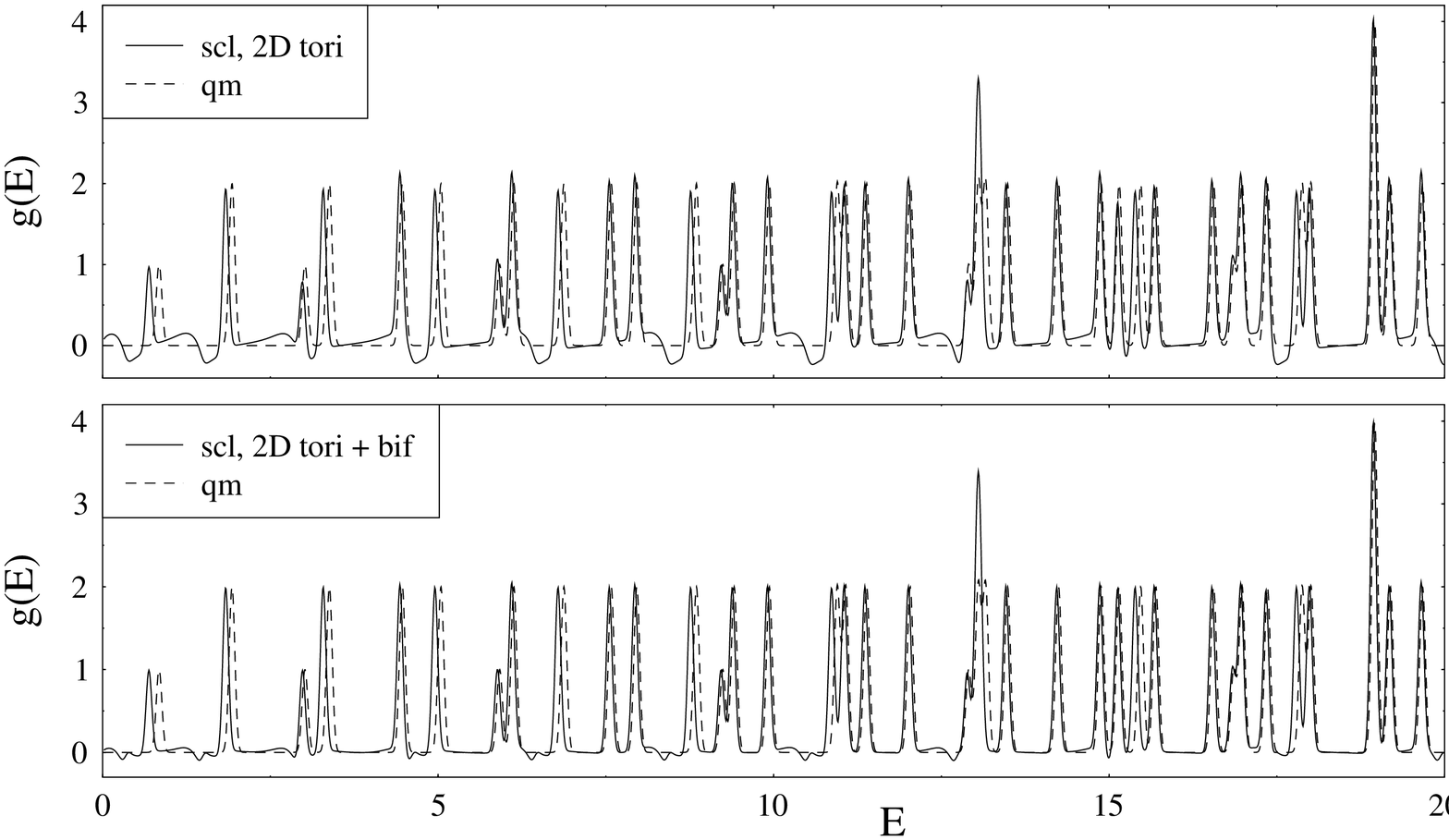}{9.2}{16.6}{
The same as \fig{dg_1}, but for the total densities of states including 
the average part \eq{gtf}, with $\gamma=0.06$. Periodic orbit sums are 
cut at $n_{max}=m_{max}=k_{max}=30$.
}

The effect of the bifurcating orbits is less dramatic when we choose a 
finer resolution of the energy spectrum, obtained with a smaller value
of $\gamma$. In \fig{dg_06} we show the total density of state including 
the average part \eq{gtf}, obtained for $\gamma=0.06$. Hereby periodic 
orbits up to $n_{max}=m_{max}=k_{max}\simeq 30$ contribute to the 
semiclassical results. We have normalized $g\,(E)$ here by the factor 
$\gamma\sqrt{\pi}$ so that the quantum result exhibits the correct 
degeneracies 1 or 2. Some apparently wrong higher degeneracies are badly 
resolved accidental (near-)degeneracies. At first glance there is not 
much of a difference between the upper and lower panels of \fig{dg_06}. 
However, a closer inspec\-tion reveals that the semiclassical degeneracies 
are wrong, with errors of up to $\sim 20\%$, in the upper part where only 
the 2D tori are included. They come much closer to the exact ones when 
including the bifurcating orbits, as seen in the lower panel. There also 
the bottom regions between the peaks are imp\-roved, the remaining small 
oscillations being numerical noise. Note that the inclusion of the 
bifurcating orbits does not affect the semiclassical peak positions which 
are exactly those of the spectrum $E_{nm}^{\rm EBK}$ given in \eq{ewkb}. 
The shifts in the peaks seen at the lowest energies are due to the typical 
errors inherent in the EBK approximation, which rapidly decrease with 
increasing energy.

\subsection{The non-integrable case $\alpha=6$}

As an example of a non-integrable situation we choose $\alpha=6$ where the 
rotating orbit R$_5$ is born at the bifurcation A$_5\to$ A$_6$ + R$_5$. In 
\fig{poinc6} we show a Poincar{\'e} surface of section $(y,p_y)$ taken at 
$x=0$. The large regular island in the middle contains the orbits A and 
R$_5$ at the origin. At its border we see an island chain of eight pairs 
of stable and unstable orbits born from a period-quadrupling bifurcation 
of the A orbit at $\alpha=5.4306$; the scaling properties of these fixed 
points were investigated in \cite{lakh}. The four drop-like small regular 
islands around $(y,p_y)=(\pm 0.431,\pm 0.362)$ contain the stable librating 
orbits S$_3$ that are born from the B orbits in the isochronous pitchfork 
bifurcation B$_3\to$ B$_2$ + S$_3$ at $\alpha=3$, as seen in \fig{q4trm}. 
[This is the same scenario as A$_3 \to$ A$_2$ + L$_3$ at $\alpha=0$, 
obtained after the transformation $\alpha\to(3-\alpha)/(1+\alpha)$ 
discussed in the introduction.] The orbits S$_3$ are still stable at 
$\alpha=6$. Since their common periods and actions are smaller than those 
of the orbit A, we expect them to influence the gross-shell structure in 
the density of states.

\Figurebb{poinc6}{-61}{37}{811}{520}{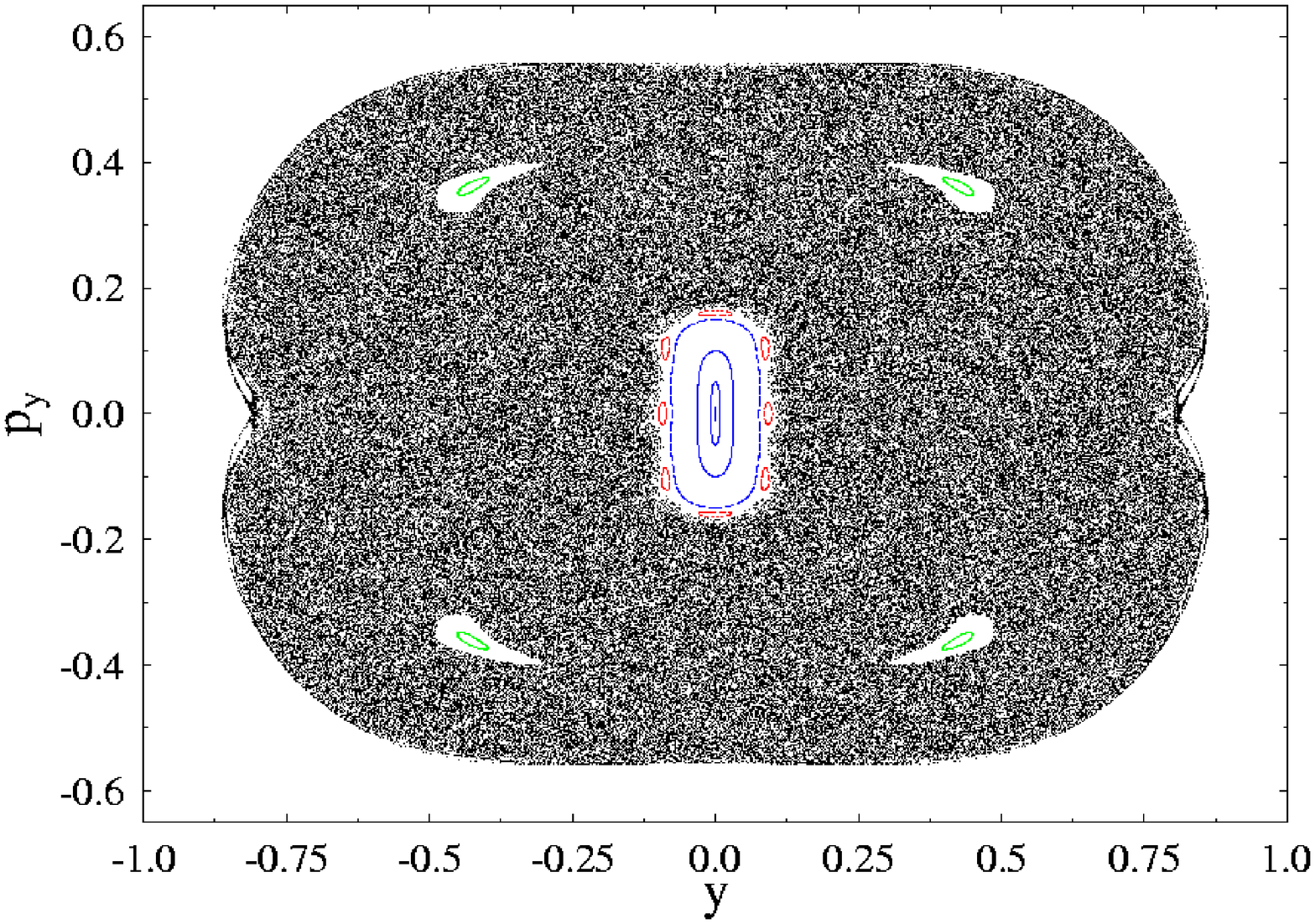}{8.6}{16.6}{
Poincar{\'e} surface of section (taken at $x=0$) for $\alpha=6$.
See text for a discussion of the visible regular islands.
}

Clearly, there is no hope of obtaining full semiclassical quantization
by summing over all periodic orbits in this mixed phase-space system, 
where many orbits of longer periods will also be close to bifurcations. 
However, we may try to obtain the coarse-grained density of states using 
only the shortest periodic orbits, as in \fig{dg_1} above, and thus
using a cut-off $k_{max}$ of the sum over the repetition number $k$. The 
contribution of the four S$_3$ orbits to the semiclassical trace formula
has the standard form for a stable isolated orbit \cite{gutz,mill}:
\be
\delta g_{sc}^{S_3}(E)=\frac{2}{\pi\hbar}\,T_{{\rm S}_3}(E)\sum_{k=1}^{k_{max}}
            \frac{(-1)^k}{\sin(k\chi_{{\rm S}_3}/2)}\,f_\gamma(kT_{{\rm S}_3})
                         \sin\left[\frac{k}{\hbar}\,S_{{\rm S}_3}(E)\right], 
                                                            \label{dgS3}
\ee
whereby the period $T_{{\rm S}_3}(E)$, the action integral $S_{{\rm 
S}_3}(E)$ and the stability angle $\chi_{{\rm S}_3}$ have been determined 
numerically. The contribution from the bifurcating A and R orbits is 
given by \eq{levbif} with $\nu_n=6$ and using \eq{an} with $\tau_4=
-11/1260$ according to \tab{t4tab} for $n=3$.

Figure \ref{dg6} shows the coarse-grained density of states obtained
with a Gaussian smoothing width of $\gamma=1.0$, including only the
$k_{max}=2$ lowest harmonics of the shortest periodic orbits. The upper
panel gives the separate contributions of the isolated S$_3$ orbits 
(solid line) and of the bifurcating A and R orbits (dotted line); the
lower panel gives their sum. The dashed lines give the quantum-mechanical 
result in both panels. As in the case shown in \fig{dg_1}, each of the
single contributions give a monotonically increasing amplitude of the
oscillations in $\delta g_{sc}(E)$, and only their superposition yields a 
beating result that reproduces the quantum-mechanical result fairly well. 
The remaining discrepancies must be due to the missing contributions 
from other orbits with longer periods, of which there are already many at 
$\alpha=6$. 

In spite of the fact that their amplitude in \eq{levbif} is larger by a 
relative factor $\hbar^{-1/4}$ than that of the isolated orbits, the 
overall contribution of the bifurcating orbits to the coarse-grained
density of states is seen in the upper part of \fig{dg6} to be lower. 
This has two reasons. First, the shortest isolated orbits $S_3$ have a 
period that is substantially smaller than that of the orbits A and R$_5$, 
which leads to a larger value of the coarse-graining factor $f_\gamma$ 
in the trace formula. Second, the orbits $S_3$ are stable and have 
therefore a stability denominator of order unity, whereas the amplitude 
in \eq{levbif} has a larger overall denominator. Still, the uniform
A + R$_5$ contribution is strong enough to cause the beating interference 
pattern seen in the total $\delta g_{sc}(E)$.

\Figurebb{dg6}{-20}{35}{795}{505}{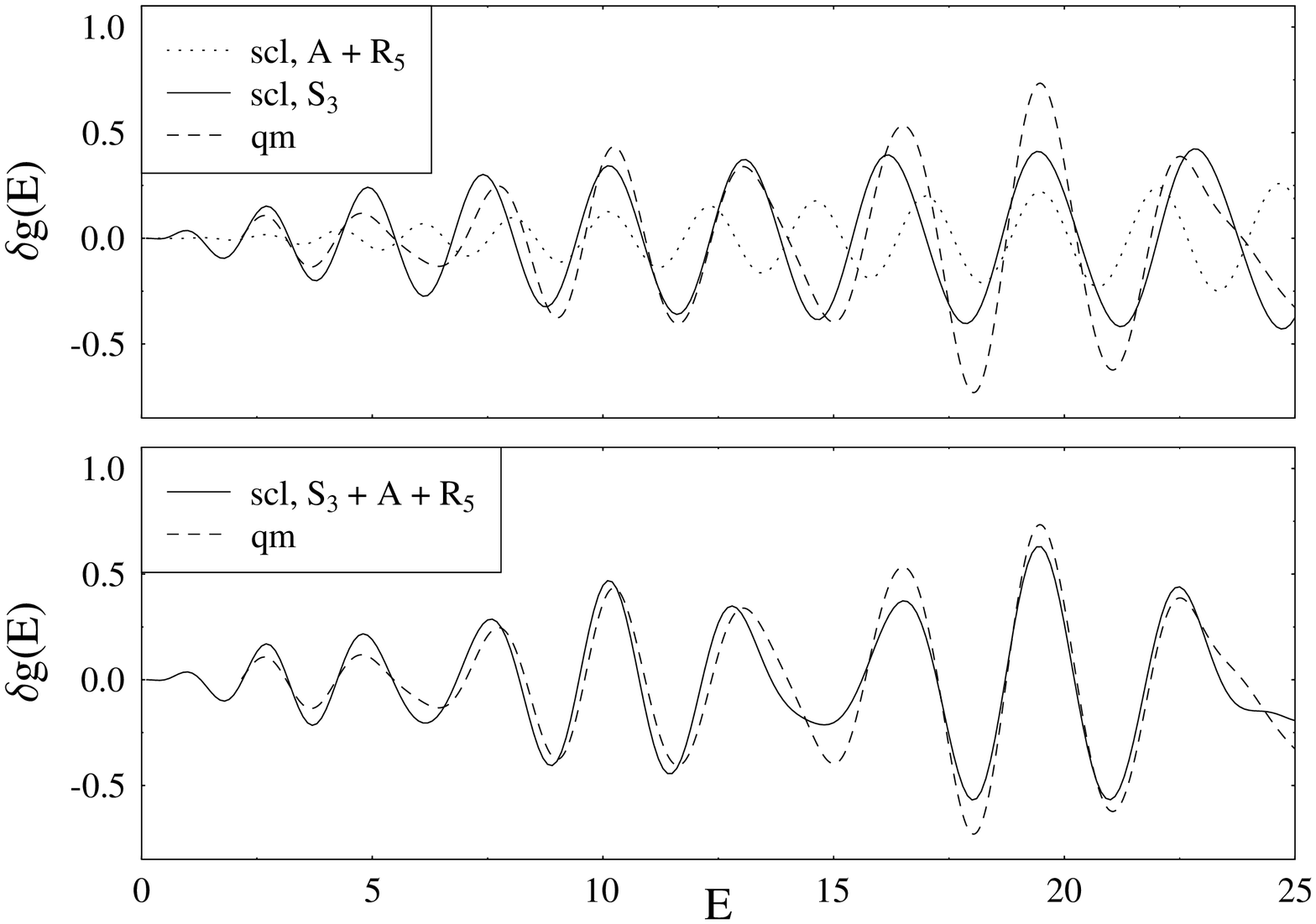}{9.25}{16.6}{
Oscillating part of density of states as in \fig{dg_1}, but at $\alpha=6$.
Upper panel: separate contributions of the stable isolated orbit S$_3$ 
(solid line) and of the bifurcating orbits A + R$_5$ (dotted line). 
Lower panel: sum of orbits S$_3$ + A + R$_5$ (solid line). Truncation 
of the trace formulae \eq{levbif} and \eq{dgS3} at $k_{max}=2$, Gaussian 
averaging width $\gamma=1.0$. The dashed line in both panels gives the 
quantum-mechanical result.
}

It is now interesting to speculate about the relative contribution of
the bifurcating orbits when the parameter $\alpha$ is further increased
on the route towards chaos. On one hand, the average stability of the 
shortest isolated orbits can be expected to decrease. On the other hand, 
\fig{uniamp} reveals that the uniform amplitude in \eq{levbif} of the 
period-one ($k=1$) orbit A with its offsprings at the bifurcation 

\Figurebb{uniamp}{-120}{70}{795}{560}{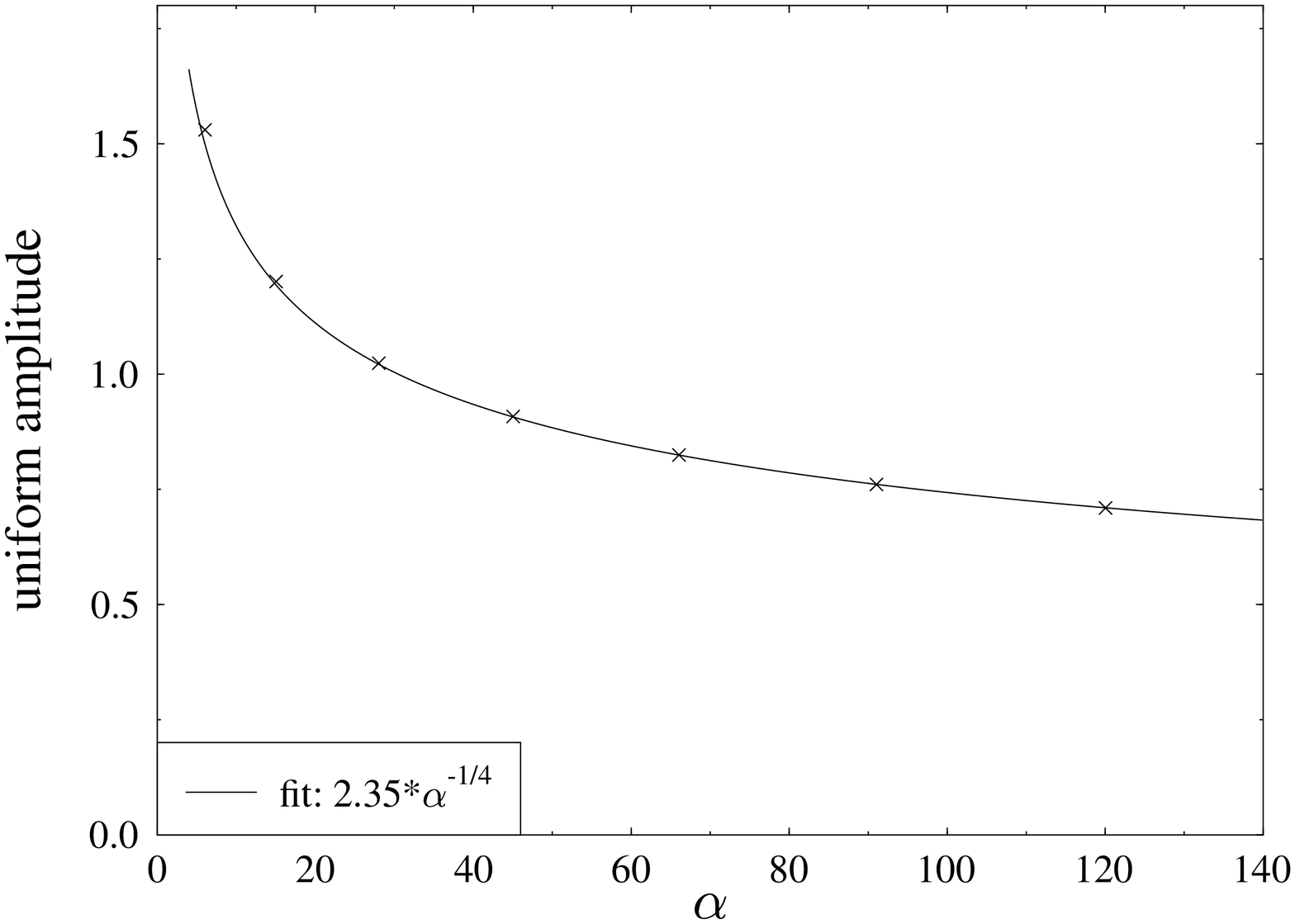}{8.3}{16.6}{
Amplitude of the uniform approximation \eq{levbif} for the bifurcating 
A orbit ($k=1$) to the semiclassical density of states ($E=1/6$). The 
crosses give the values obtained using the values of $\tau_4$ in 
\tab{t4tab} for $n>0$ at the bifurcation points $\alpha_n$; the solid 
line is an asymptotic fit for large $\alpha$.
}

\noindent
points also decreases with increasing $\alpha_n$. (The crosses give the 
analytical values obtained using the $\tau_4$ values from \tab{t4tab},
and the solid line is a numerical fit showing that the amplitude decays 
asymptotically like $\alpha^{-1/4}$.) Taking furthermore account of the 
fact that more and more new pairs of isolated orbits are created from 
the bifurcations along the route to chaos (and new orbits bifurcate from 
these again), even the number of shortest (period-one) isolated orbits 
increases very fast. We therefore tend to conclude that the relative
weight of the bifurcating orbits in the coarse-grained density of 
states will decrease with $\alpha$. It would, however, require a major 
numerical effort to study this balance in more detail and, in particular, 
to investigate the contributions from higher repetitions to the finer 
details of the density of states.

\section{Summary}

We have formulated a perturbative scheme to calculate analytically the 
shapes of periodic orbits bifurcating from the straight-line orbits in 
the quartic oscillator potential, exploiting the nice properties of the 
periodic Lam{\'e} functions in terms of Jacobi elliptic functions. In 
order to simplify the recursive determination of the unknown constants
in the perturbation expansion, we have used an unconventional {\it a 
posteriori} rescaling of the perturbed solutions that takes care of the 
varying periods of the bifurcated orbits.

For stable period-one orbits we are able to give analytical expressions 
of the resulting perturbative series. Even when the perturbation 
parameter $\eps$ is greater than unity, satisfactory convergence 
to the numerically obtained solutions can be reached by going up to 
order $\eps^4$ where the necessary algebraic work is still not too 
demanding.

We have studied semiclassical trace formulae for approximating the 
quantum-mechanical density of states of the quartic oscillator. In the 
separable case $\alpha=0$ we could give a completely analytical trace
formula, consisting of the contributions \eq{gtf}, \eq{dgsc2} and 
\eq{dgsc1}. It was shown numerically to give a very good approximation 
to the exact quantum-mechanical density of states. Its low-frequency 
part, extracted with an energy coarse-graining parameter $\gamma=1$, 
reveals a strong quantum beat. This beat can only be reproduced 
semiclassically when the shortest periodic orbit families are allowed 
to interfere with the isolated orbits A and L$_3$ taking part in a 
pitchfork bifurcation. The contribution of the latter was calculated 
using a slightly modified uniform approximation for generic pitchfork 
bifurcations given in \eq{dgsc1}. The high-resolution spectrum is 
dominated by the Berry-Tabor part \eq{dgsc2} which yields exactly the 
peaks corresponding to the EBK spectrum. The quantum degeneracies are, 
however, substantially affected by the contributions from the 
bifurcating orbits. 

In the non-integrable situation $\alpha=6$, which exhibits a strongly 
mixed phase space, we could also reproduce the coarse-grained quantum 
density of states semiclassically, including besides the bifurcating 
period-one orbits A and R$_5$ also the shortest stable isolated orbits 
S$_3$ which had to be obtained numerically. The two types of orbits 
interfere, again leading to a beat structure. Finally, we have shown 
the uniform contribution of the bifurcating A orbits and its offsprings 
to decrease like $\alpha^{-1/4}$ for further increasing values of the 
chaoticity parameter $\alpha$.

With these results we have demonstrated the possibility of accessing 
analytically the periodic orbits in a system with mixed classical 
dynamics. We hope to stimulate further research in this direction,
in particular towards semiclassical studies of the role of periodic 
orbits and their bifurcations in connection with spectral statistics 
\cite{q4ls,spec}, which would have exceeded the scope of our present 
short report. We also aim at further improving our understanding of 
the uniformization of semiclassical trace formulae, including  
bifurcations of higher codimensions and -- as a long-term (and perhaps 
too ambitious?) goal -- including complete bifurcation cascades.

\newpage

\noindent
{\Large \bf Note added in proof}

\bs

We have confirmed the result \eq{dgsc1} in an alternative way,
without making use of any normal form. We start from the Poisson 
summation-integration of the density of states 
$\sum_{n_x,n_y}\delta(E-E_{n_x,n_y}^{EBK})$ using the 
EBK spectrum \eq{ewkb}, like in the standard derivation \cite{beta}
of \eq{dgsc2}. The first integration can be done exactly exploiting 
the delta function. For the edge contributions corresponding to the 
A and L$_3$ orbits, the usual stationary phase approximation cannot 
be used and the second integration must be done more carefully. Its 
asymptotic evaluation, valid here for $k\SA/\hbar\simg 2$, leads 
precisely to the analytical result \eq{dgsc1}. 

\bs
\bs

\noindent
{\Large \bf Acknowledgements}

\bs

\noindent
We are grateful to J{\"o}rg Kaidel for helpful discussions and Christian 
Amann for providing us with a program for the numerical computation of 
the quantum spectrum. SNF, AGM and MM acknowledge the hospitality of the
University of Regensburg during their research visits and financial 
support by the Deutsche Forschungsgemeinschaft.

\end{document}